\documentclass{article}

\usepackage{amsmath}

\usepackage[dblblindworkshop, final]{neurips_2025}
\workshoptitle{Data on the Brain \& Mind}

\bibpunct{(}{)}{;}{a}{,}{,}

\usepackage[utf8]{inputenc} 
\usepackage[T1]{fontenc}    
\usepackage{url}            
\usepackage{booktabs}       
\usepackage{amsfonts}       
\usepackage{nicefrac}       
\usepackage{microtype}      
\usepackage{xcolor}         
\usepackage{graphicx}
\usepackage{appendix}
\usepackage{subfigure}
\DeclareMathOperator*{\argmax}{arg\,max}

\usepackage{placeins} 

\usepackage{hyperref}       

\title{Manifolds and Modules:\\How Function Develops in a Neural Foundation Model}

\author{%
  Johannes Bertram \\
  University of Tübingen \\
  \texttt{johannes.bertram@student.uni-tuebingen.de}
  \And
  Luciano Dyballa \\
  School of Science \& Technology \\
  IE University \\ 
  \And
  T. Anderson Keller \\
  The Kempner Institute for Natural and Artificial Intelligence\\
  Harvard University \\ 
  \And
  Savik Kinger \\
  Department of Computer Science\\
  Yale University\\
  \And
  Steven W. Zucker\\
  Depts. of Computer Science and Biomedical Engineering\\
  Wu Tsai Institute\\
  Yale University\\
}

\begin{document}

\maketitle

\begin{abstract}
  Foundation models have shown remarkable success in fitting biological visual systems; however, their black-box nature inherently limits their utility for understanding brain function. Here, we peek inside a SOTA foundation model of neural activity \citep{wang_foundation_2025} as a physiologist might, characterizing each `neuron' based on its temporal response properties to parametric stimuli. We analyze how different stimuli are represented in neural activity space by building \textit{decoding manifolds}, and we analyze how different neurons are represented in stimulus-response space by building \textit{neural encoding manifolds}. We find that the different processing stages of the model (i.e., the feedforward \textit{encoder}, \textit{recurrent}, and \textit{readout} modules) each exhibit qualitatively different representational structures in these manifolds. The \textit{recurrent} module shows a jump in capabilities over the \textit{encoder} module by ``pushing apart'' the representations of different temporal stimulus patterns; while the \textit{readout} module achieves biological fidelity by using numerous specialized feature maps rather than biologically plausible mechanisms. Overall, we present this work as a study of the inner workings of a prominent neural foundation model, gaining insights into the biological relevance of its internals through the novel analysis of its neurons' joint temporal response patterns.
\end{abstract}

\section{Introduction}
\label{introduction}
\begin{figure}[t]
  \centering
  \includegraphics[width=\textwidth]{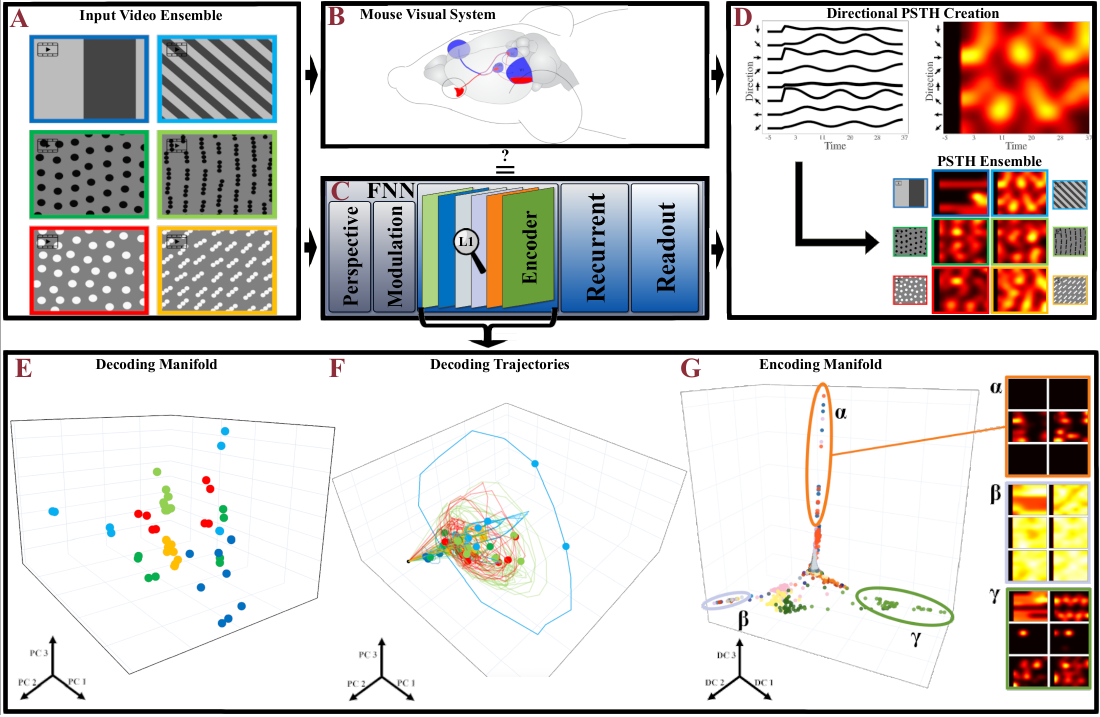}
  \caption{\textbf{Set up and techniques.} \textbf{A}: The stimuli consist of drifting gratings (at two spatial frequencies and 8 directions of motion) plus dotted and oriented flows, at two contrasts, drifting in 8 directions.
  \textbf{B, C}: Stimuli exercise both the mouse visual system (data from published literature, image from \citet{Wilks13}) and the trained FNN (this paper) to yield activity (firing rate) in time for each direction. \textbf{D}: The firing rates are collected as a PeriStimulus Time Histogram (PSTH), denoted as a heatmap image (higher firing rate is brighter). \textbf{E}: Neural decoding manifold (each point is a trial; coordinates are PCA-reduced neural firings); colors for each trial point match the boxes around stimuli in \textbf{A}. While the trials are weakly clustered by stimulus, the representations do not allow for clear classification at this stage. \textbf{F}: Decoding trajectories show development of neural activity over time for each stimulus, also in PCA coordinates. As expected in the early stage feedforward \textit{encoder}, neural activity barely changes after the onset of activity compared to the 0-activity point (black). Only circular temporal developments are observable for periodic input stimuli, such as moving gratings (light blue). \textbf{G}: Neural encoding manifold, in which each point is a neuron, in diffusion coordinates. Average PSTHs for neurons in circled clusters show average activity for each of the stimulus classes (arranged as in \textbf{A}). Note the multi-selectivity of neurons to different stimulus classes and especially the ``amplification'' induced by neurons in cluster $\beta$. We study the \textit{encoder} (shown above), \textit{recurrent} and \textit{readout} modules, and ask whether they have analogues in the mouse visual system.
    }
  \label{fig1}
  \vspace{-2mm}
\end{figure}
 
Deep neural network models are powerful tools for modeling the mouse visual system by learning to predict neural responses directly from visual input \citep{cowley_compact_2023, ustyuzhaninov_digital_2022, huang_deep_2023}.
While much prior work has explored different computational models  \citep{averbeck_neural_2006, ustyuzhaninov_digital_2022, qazi_mice_2025}, foundation models are becoming extremely valuable: they not only fit neural activity at the unit level but are capable of generalizing beyond training data \citep{wang_foundation_2025, li_v1t_2023}. Nevertheless, their complexity can overwhelm understanding of the learned input-output map. Control theory teaches us that, without a perfect model, one must ``look inside the box'' to achieve identifiability (cf. \citep{aastrom2012introduction}). Without this, we cannot guarantee correct, robust, and generalizable behavior, especially on out-of-distribution data, to confidently build hypotheses about the brain using these models. Thus, we study the state-of-the-art foundation model, the FNN \citep{wang_foundation_2025}, as a (computational) neuroscientist might. The FNN consists of multiple stages (Figure~\ref{fig1}C) and was trained on the MICrONS functional connectomics dataset \citep{bae_functional_2025} using artificial and `natural' videos across multiple animals. We use modeling tools available online (references in Methods), stimuli similar to those used in FNN's original training \citep{wang_foundation_2025}, and add naturalistic flow stimuli used in mouse physiology \citep{dyballa18flow}. The last of these allows us to examine out-of-distribution performance (Figure~\ref{fig1}A). 

\looseness=-1
Prior work has explored how artificial models represent neural responses \citep{averbeck_neural_2006, ustyuzhaninov_digital_2022, qazi_mice_2025}, and has examined deep neural networks with regard to functionality; some are supportive \citep{dnns, yamins2014ctxdnn, margalit_unifying_2024}, while some raise questions \citep{serre2019deep}, in many different species. The FNN is a data-driven predictive model with a Gaussian readout \citep{klindt2018neuralidentificationlargepopulations, turishcheva2024reproducibilitypredictivenetworksmouse, nellen2025learningclusterneuronalfunction, lurz2021generalization} that interprets the output as per-neuron basis functions with individual readout weights. The readout provides an encoding embedding of biological neurons. For comparability, we use our encoding method to compare the embeddings of biological neurons and individual readout neurons, investigating both the final and the readout embeddings. Different loss functions have been used for fitting mouse models \citep{nayebi_mouse_2023, bakhtiari2021functional, shi_mousenet_2022}, and others have studied decoding manifolds for mouse \citep{froudarakis2020object, beshkov2022geodesic, beshkov2024topological}, focusing on topological properties. For a recent general review, see \citep{doerig2023neuroconnectionist}.

\looseness=-1
To understand how the FNN performs, we analyze the internal representations at each processing stage (Figure~\ref{fig1}) using three techniques popular in neuroscience. (1) We build \textit{neural decoding manifolds} \citep{chung_neural_2021}, in which trials are embedded in the space of neural activity coordinates (Figure~\ref{fig1}E), then dimensionality-reduced using PCA \citep{cunningham2014dimensionality}. Typically, trials involving the same stimulus cluster together, facilitating a read-out of the brain's state. (2) To switch from trials to neurons,  we build \textit{neural encoding manifolds} (Figure~\ref{fig1}G) \citep{dyballa_population_2024} in which each point is a neuron in the space of stimulus-response coordinates, dimensionality-reduced using tensor factorization \citep{williams_unsupervised_2018}. Proximity between neurons in an encoding manifold denotes similar responses to similar stimuli; i.e., groupings of neurons that are likely to share circuit properties. Finally, (3) the relationship between these two manifolds is captured by the temporal evolution of each neuron's activity for each stimulus trial. We note the popular view that a `neural computation' can be viewed as the result of a dynamical system in neural state space \citet{hopfield1984neurons}. We plot these both as PSTHs (Fig.~1.D.) and as streamline traces (decoding trajectories, Fig.~1.F.). While streamline representations have been used previously for decision tasks \citep{duncker_dynamics_2021} and the motor system \citep{churchland2012neural, gallego}, we note: (i) the activity integral along such \textit{decoding trajectories} (Figure~\ref{fig1}F) defines the decoding manifold, while (ii) shared tubular neighborhoods specify position in the encoding manifold.These three perspectives allow us to examine distinct aspects of alignment: (1) Decoding manifolds indicate whether the model preserves stimulus separability observed in biological systems; (2) Encoding manifolds assess whether the functional topology of neurons resembles that of the brain; (3) Trajectories evaluate whether the model executes computations using dynamics similar to those in the brain. Importantly, a model may demonstrate alignment at one level while failing at others. To our knowledge, this is the first time all three of these techniques have been utilized together for analysis of a perceptual system; i.e., toward interpretability for a foundation model. For a review of classical encoding/decoding in neuroscience, see \citep{mathis2024decoding}.

With this framework, we ask: \textit{Do neural decoding and encoding manifolds reveal new insights into how foundation models represent temporal response patterns? Are the representations brain-like?} We hypothesize that each processing stage contributes distinct representational capabilities, all essential for fitting neural data. In particular, one might expect the \textit{recurrent} module to enrich the temporal structure of representations, analogous to cortex, and the encoder layers to resemble a retina with relatively limited recurrence.

\section{Methods}
\label{methods}

\looseness=-1
Our work makes novel use of publicly available open-source resources. Specifically, we employed the pretrained foundation model of neural activity (denoted FNN) provided by \citet{wang_foundation_2025}, available \href{https://github.com/cajal/fnn/tree/main}{here}; and the stimulus generation tools and neural encoding manifold construction pipeline introduced by \citet{dyballa_population_2024}, accessible \href{https://github.com/dyballa/NeuralEncodingManifolds}{here}. Below we briefly outline our methods, and refer readers to Appendix \ref{app:methods} for the full details.

\paragraph{Model:} The FNN consists of five modules: perspective, modulation, \textit{encoder}, \textit{recurrent}, and \textit{readout}. The perspective and modulation modules model the mouse’s state and transform the inputs to approximate the actual visual information received.
Thus, only the \textit{encoder}, \textit{recurrent}, and \textit{readout} modules perform the core computation and are the focus of this work. The \textit{encoder} module is a 10-layer DenseNet-style convolutional encoder. Notably, it includes 3D convolutions, which in principle enable the encoder to capture temporal patterns for up to 12 time steps. The \textit{recurrent} module is preceded by an attention layer and consists of a convolutional LSTM, followed by a single convolutional layer that produces its output.  
This feedforward–recurrent combination constitutes the core of the FNN, which is trained on data from all mice in combination. Finally, a separate \textit{readout} module is trained on each mouse individually: it performs an interpolation on the recurrent output followed by a linear transformation to produce the FNN output.

\paragraph{Stimuli:} Our stimulus set is composed of drifting square-wave gratings and optical flows moving in eight directions. The flow stimuli include oriented (lines) and non-oriented (dots) stimuli with spatial frequencies between 0.04 and 0.5 \(\frac{\text{cycles}}{\text{deg}}\). This yields 88 unique input sequences with stochastic initial positions and velocities. The stimuli are scaled and cropped to match the FNN requirements. A subset of these stimuli are visualized in Figure \ref{fig1}A. To ensure that these stimuli would drive the network in a representative manner, we compared the output of the network for these stimuli with the output for the original natural movie stimuli used to train the network (Appendix Figures \ref{fig7}, \ref{osi}); and found the results to be quantitatively similar in all measured respects.

\paragraph{PSTH visualization:} To visualize the network responses to many stimuli concisely, we group together the model's PeriStimulus Time Histogram responses (PSTH) corresponding to all flow directions of a single given stimulus pattern, forming the black-yellow heatmaps of spike rate, with time on the x-axis and flow direction on the y-axis, depicted in Figure \ref{fig1}D. We then organize these joint PSTH plots according to 6 of the unique stimulus patterns in a 2$\times$3 grid.

\paragraph{Decoding manifolds \& Trajectories:} Following traditional analysis techniques, we first constructed decoding manifolds by performing PCA on the stimulus-time-averaged activity data. In total therefore, the decoding manifold contains 48 points, one for each unique sequence, and colored by the corresponding base-stimulus (shown in Figure~\ref{fig1}). Different spatial frequencies of the same stimulus are summarized with the same color. To construct \textit{decoding trajectories}, we treat each time step as a separate data point rather than averaging across time before applying PCA. We compare with biological \textit{decoding trajectories} using the experimental data from \citet{dyballa_population_2024}. For additional quantitative analysis of these trajectories, we compute a novel `clustering' of bundles of trajectories, which we term \emph{tubularity}, yielding values for `Tightness' and `Crossings', depicted in Figure \ref{fig:tubes} (definition in Appendix~\ref{tubular:methods}). 

\paragraph{Encoding manifolds:} To understand the response properties of \emph{neurons} with respect to all stimuli (rather than the representation of \emph{stimuli} in the space of all neurons), we finally construct \emph{encoding manifolds}. At a high level, these manifolds allow one to examine the global topology of neuronal populations based on their stimulus selectivities and temporal response patterns \citep{dyballa_population_2024}. The neural encoding manifold is constructed in a three-step procedure. First, a 3-tensor is built with the temporal responses from each neuron for each stimulus, and decomposed using Nonnegative Tensor Factorization (details in Appendix); each component is comprised of neural, stimulus, and temporal response factors. The neural factors then serve as position coordinates, embedding the neurons into a stimulus-response framework called the neural encoding space. Second, we construct a data graph in this neural encoding space using the IAN algorithm \citep{dyballa_ian_2023}. Third, applying diffusion maps \citep{coifman_geometric_2005, coifman_diffusion_2006} to the data graph yields the manifold. We follow the methodological choices of \citet{dyballa_population_2024}, where extensive parameter analysis for biological neural data was conducted.

\begin{table}[b]
\vspace{-2mm}
  \caption{Stimulus classification accuracy for Leave-One-Out 3-Nearest Neighbor (3-NN) and Logistic Regression (LR) classifiers trained on each layer's activations. Methods' details in Appendix \ref{app:methods}.}
  \label{tab1}
  \centering
  \begin{tabular}{lllllllllll}
    \toprule
        Accuracy & Enc1 & Enc2 & Enc4 & Enc5 & Enc7 & Enc8 & Rec & RecOut & Readout & Out \\
    \midrule
     LR & 0.59 & 0.62 & 0.66 & 0.65 & 0.71 & 0.74 & 0.89 & \textbf{0.90} & 0.88 & 0.77 \\
     3-NN & 0.41 & 0.66 & 0.58 & 0.52 & 0.53 & 0.61 & \textbf{0.73} & 0.64 & 0.63 & 0.67 \\
    \bottomrule
  \end{tabular}
\end{table}

\begin{figure}[h]
  \centering
  \includegraphics[width=\textwidth]{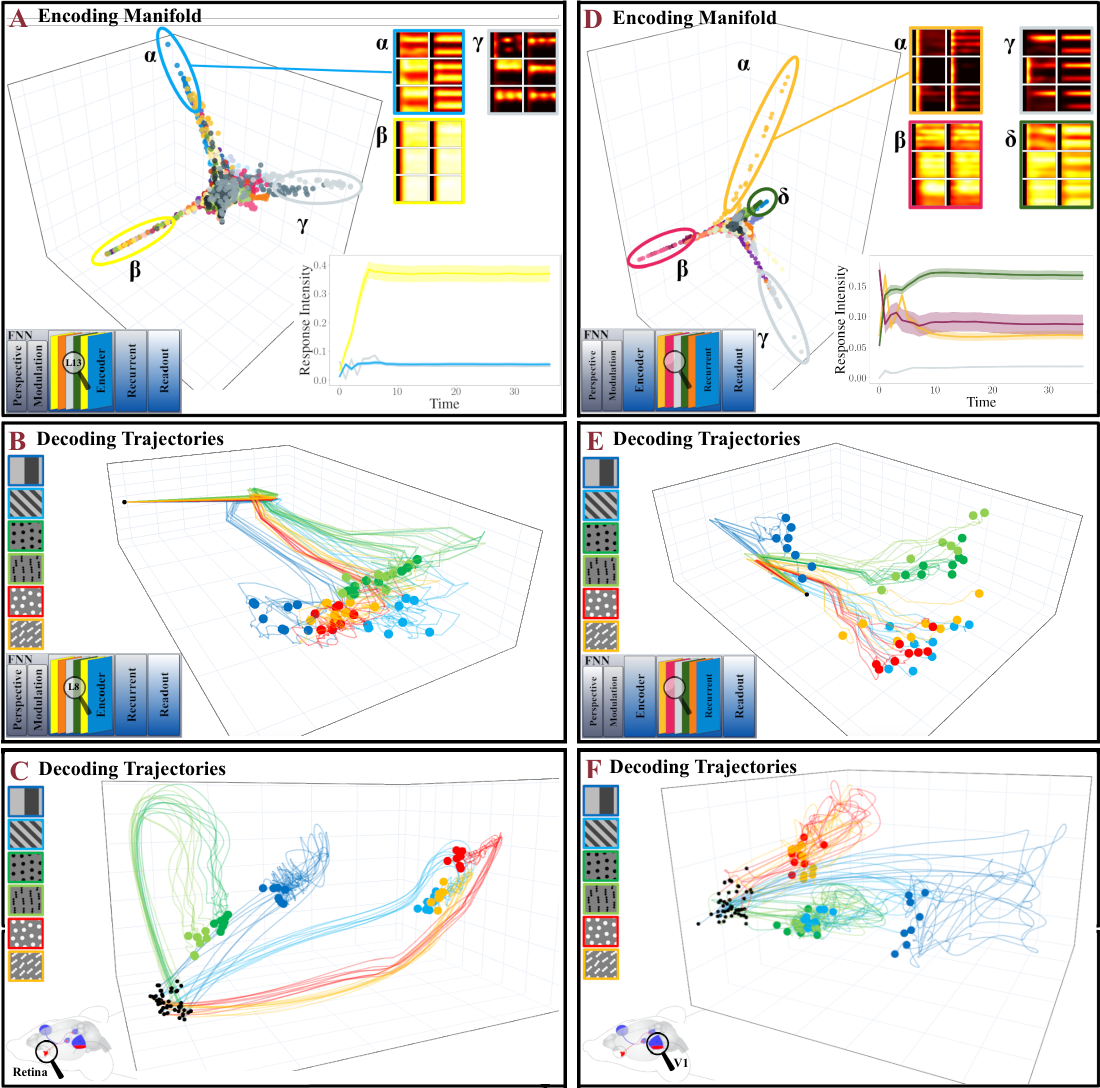}
  \caption{\textbf{Encoding and recurrent layers}.
  \textbf{A}: Encoding manifold for final encoding block. PSTHs for arm $\beta$ amplify all stimulus signals; inset shows mean response intensity development $\pm1$ s.e.m. within units in $\beta$ (yellow) compared with others.
  \textbf{B}: Explosive growth of trajectories for FNN is caused by initial intensity increase in $\beta$. Ensuing temporal dynamics are negligible. This differs from the trajectory bundles found in mouse retina (\textbf{C}), showing stimulus dependence instead of nonselective intensity induced temporal patterns. 
  \textbf{D}: Recurrent hidden state shows multi-selectivity of units and no explosive intensity growth (cf. inset). \textbf{E}: Decoding trajectories show increased stimulus-dependent temporal patterns leading to better discriminability of stimuli in PCA space. However, trajectories are more temporally monotonic than in primary visual cortex (\textbf{F}).
  }
  \label{fig2}
  \vspace{-1mm}
\end{figure}

\begin{figure}[h]
  \centering
  \includegraphics[width=\textwidth]{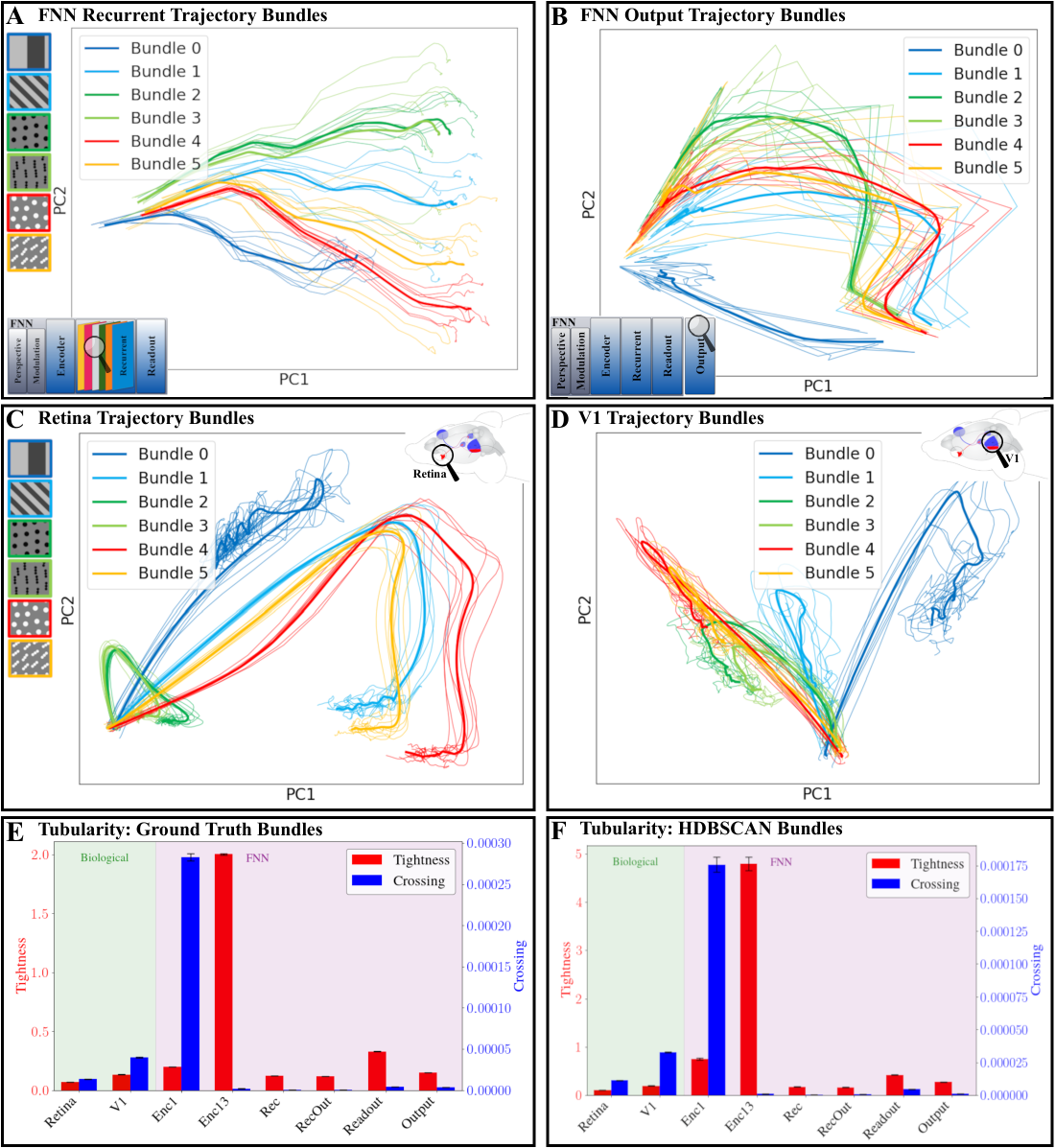}
  \caption{{\bf Tubularity of \textit{recurrent} module and \textit{output} compared to mouse data.} Similar decoding trajectories can be clustered into bundles and averaged. Here we cluster by stimulus class; the mean contour is displayed in the stimulus-class color for {\bf A.} the recurrent module (note class separation developing) and {\bf B.} the output layer (note the `circular' dynamics). These differ from biological trajectories for {\bf C.} mouse retina and {\bf D.} V1. These differences are quantified by the tightness and crossing measures (Appendix~\ref{tubular:methods}) for both ground truth and unsupervised (HDBSCAN) groupings.
  }
  \label{fig:tubes}
  \vspace{-0mm}
\end{figure}


\section{Results}

We built encoding manifolds, as well as decoding trajectories and manifolds, for all layers (of considered modules) in the FNN. Here we highlight the results most helpful toward interpreting the computational role of each stage of the FNN network.
Beginning with the first encoder layer (L1), we found that its decoding manifold was poorly clustered (Figure~\ref{fig1}E), with the different stimulus classes quite mixed. This implies that, at this point within the FNN, the latent feature representation is not sufficient to distinguish between the different stimuli (indeed, its classification accuracy is lowest; see {Table~\ref{tab1}). The decoding trajectories for L1, however, reveal a more complete picture: from Figure~\ref{fig1}F we note that periodic stimuli are represented as loops, likely due to the translation equivariance of the convolutional layers of the encoder preserving the circular geometric structure of these stimulus sequences \citep{pmlr-v48-cohenc16}. However, we see that these loops can take on many different forms 
(such as the high spatial frequency gratings, shown in light blue), defined by the responses of individual neurons to each stimulus. 
Finally, the encoding manifold for L1 (Figure~\ref{fig1}G) completes the characterization by revealing that most neurons belonging to the same feature map (points with the same color label) form contiguous clusters, or regions, over the manifold; this is not entirely surprising given the weight-sharing property of these convolutional layers. Nevertheless, several feature maps are found mixed into the same ``arm'' (labeled $\beta$). Examining the response profile (PSTH ensemble) of these neurons in detail, we notice strong, continuous activity throughout the trial duration to all stimulus classes.

We now move on to the late-stage encoder, layer 8 (L8). First, although its encoding manifold shows that the grouping by feature maps is still apparent, especially in the right-hand side of the manifold (Figure~\ref{fig2}A), the overall manifold appears less clustered and more mixed. On the other hand, again we find a poorly-selective ``intensity arm'' of neurons ($\beta$) from multiple feature maps representing a strong response to all stimuli. This is supported by plotting the mean activity of neuron groups (inset in Figure~\ref{fig2}A). The marked increase in response magnitude early in the trial among units in the $\beta$ arm can be readily noted in the decoding trajectories' visualization (Figure~\ref{fig2}B}). Further investigation revealed that the intensity arm arises from padding artifacts at the edges of feature maps. Sampling from the feature maps' central regions eliminates the intensity arm and the shared activity development in the decoding trajectories (see Supplemental Figure~\ref{fig:no_intensity}). However, these artifacts impact the representation, as the smoothness of the intensity arm visualizes the spread of the information of the intensity artifacts across the feature maps. Since these artifacts are present during normal network operation, excluding them would misrepresent the model's actual internal dynamics. We therefore retain these artifacts in our manifold analysis. The L8's decoding manifold was qualitatively similar to Layer 1's (not shown).

How do these findings for the encoder stage (L1 and L8) compare to the retina, the first stage in the mouse visual system \citet{baden2016}? We applied the same procedure to analyze the physiological data from \citet{dyballa_population_2024}. The non-selective groups of neurons with high activity ($\beta$ arms in Figs.~\ref{fig1}E and \ref{fig2}A) are the first departure from what is found in biological networks: in the retina there are no such non-selective neurons. Such low-selectivity in cortex is restricted to inhibitory (inter)neurons, and continuously mixes with other, more selective responses; they do not segregate as an arm, or cluster.
Perhaps the biggest difference is that retinal decoding trajectories formed largely segregated, stimulus-dependent bundles whose temporal dynamics allowed for linear separability during much of the trial's time-course (Figure~\ref{fig2}C,F). As will be shown, this is also in stark contrast with what was found along the encoder stage of the FNN (Figure~\ref{fig2}B). Thus, despite temporal convolutions, the FNN feedforward encoder appears to lack biologically plausible stimulus-dependent temporal patterns and mainly reports features present in the input with varying intensity.

The \textit{recurrent} module is qualitatively different. Its encoding manifold shows that different regions exhibit a  variety of distinct selectivity and temporal response patterns, cf. their PSTHs (Figure~\ref{fig2}D). Furthermore, although the segregation by feature map is still present, no longer do we find a cluster of neurons with no selectivity (e.g., even the highlighted $\beta$ and $\delta$ groups show clear selectivity for particular directions or orientations). Moreover, this is the first stage where the FNN is capable of reasonably decoding the different stimulus classes, as revealed by the somewhat segregated bundles of decoding trajectories in Figure~\ref{fig2}E. In fact, this is where the network reaches its highest stimulus classification accuracy (see Table~\ref{tab1}). 

While the recurrent module shows the presence of stimulus-dependent temporal patterns, the organization of decoding trajectories is noticeably more entangled than both retina and V1 (compare with Figure~\ref{fig2}C,F). This phenomenon is quantified using \emph{tubularity} metrics based on the geometry of the observed decoding bundles (see Method). We found that tubes have similar tightness from the recurrent stage onwards, but fall short of representing all complexity of activity development in biology as shown by significantly lower crossings (p<0.005, Bonferroni-corrected, for all layers) (Figure~\ref{fig:tubes}). 

The final stages of the network---the \textit{readout} and \textit{output} layers---are different again. The encoding manifold for the readout layer is highly disconnected (Figure~\ref{fig3}A), with each cluster corresponding almost exclusively to neurons sampled from a single feature map. Each feature map exhibits a distinct response pattern that is invariant across neurons within it. Compared to this, the biological results (e.g., \citet{baden2016, dyballa_population_2024}) show more variability within cell ``types'', even in the retina. Curiously, and despite this intra-map constancy, the large number of feature maps (see PSTHs) and the rich dynamics within each one, somehow enable the \textit{output} to represent the complex behavior of neurons (Figure~\ref{fig3}B). These behaviors are captured in the FNN output via a linear combination of \textit{readout} features. Since classification accuracy has declined slightly at this stage, but orientation and direction selectivity agree (Supplemental Figure~\ref{fig:no_intensity}), we conjecture these dynamics are interpolating the spiking activity. 

\vspace{0.5cm}
\section{Discussion}

Decoding manifolds (and trajectories) allow us to compare whether networks can achieve similar degrees of stimulus representation and separability. Encoding manifolds, on the other hand, allow us to check at a global level how the responses of individual neurons (and their global organization) compare to those of biological neurons; in other words, whether the FNN and biological networks employ similar encoding mechanisms for achieving similar outputs. Since decoding trajectories are a surrogate for ``computation'' as dynamics over neural state space (cf. \citep{hopfield1984neurons}, this investigation moves beyond pairwise or average unit comparisons (e.g., RSA \citep{kriegeskorte2008representational}) and may be useful in analyzing other foundation models.

Our analysis of the FNN revealed an increasing richness of representation up to the \textit{recurrent} module (cf. \citet{hoeller2024bridging} and contrasting with \citet{xu2023multimodal, nayebi_mouse_2023, froudarakis2020object}), albeit with most PSTHs revealing a lack of typical biological-like responses \cite{ringach2016, ko2011}.  Since the FNN was trained to predict neural spike trains, classification evolved implicitly (cf. {Table~\ref{tab1})). Thus it is likely that the recurrent features are sufficiently complex for feature representation and that the subsequent modules work toward fitting the neural data instead.

However, the highly clustered topology of the latent representation found for the \textit{readout} module was not a good fit for  retina or cortex (cf. \citet{baden2016, dyballa_population_2024}, nor for higher visual areas (cf. \citet{glickfeld2017higher, dyballa2024functional, yu2022selective}). Regardless, the rich dynamics within each feature map (see PSTHs), combined with the large number of them, seem to enable the \textit{output} layer to represent the complex behavior of neurons (Figure~\ref{fig3}B), resulting in the network's high performance in predicting neural activity. Still, it is somewhat surprising that such behaviors are produced in the FNN output via a simple linear combination of \textit{readout} features---one would expect that fitting the neural activity should happen throughout the entire network, and not as a separate appendage module.

\begin{figure}[h]
  \centering
  \includegraphics[width=\textwidth]{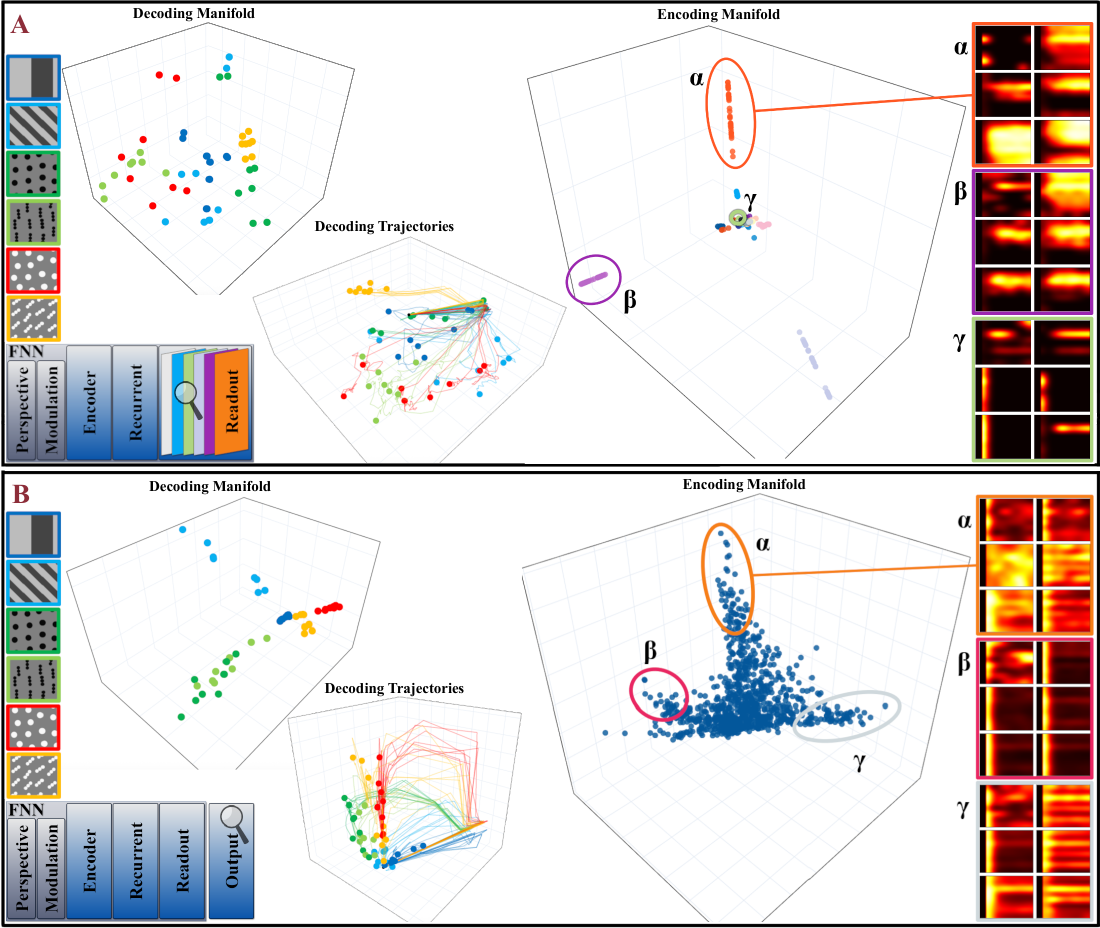}
  \caption{\textbf{Contrasting the readout (A) and output (B) layers.} While the decoding manifolds and trajectories appear qualitatively similar, the encoding manifolds have remarkably distinct topologies: while the \textit{readout} module is highly clustered, the \textit{output} is continuous. The clustered topology is caused by the interpolation step producing a large amount of features with low within-feature variability. The smooth output is obtained by collapsing the many feature maps to a single output by a linear combination.
  }
  \label{fig3}
  \vspace{-0mm}
\end{figure}

\vspace{0.2cm}
\paragraph{Limitations}
Our analysis utilized a single foundation model, due to the limited availability of other video-based foundation models of neural activity over time. Moreover, we worked with a restricted set of stimuli (see Methods) to ensure comparability to biological results. Nevertheless, there is evidence that these stimuli exercise much of the mouse visual cortex \citet{dyballa18flow}, so they provide at least a necessary component for out-of-sample examination. Moreover, we show that these stimuli exercise FNN like the natural movies on which they were trained (Appendix Figure~\ref{fig7}), empirically validating their usefulness. 

\section{Conclusion}
\label{conclusion}

We found a rich diversity of encoding and decoding topologies in the FNN, highlighting its capability to fit complex neural data. Distinct representations emerge from each module, reflecting its architecture: First, the \textit{recurrent} module appears to learn generalizable representations of temporal stimuli, encouraging uniformity and alignment, as in general self-supervised foundation models \citep{wang2022understandingcontrastiverepresentationlearning}. Second, we found that the \textit{readout} module accounts for rich biological variability, but does so by relying on a large number of self-similar feature maps, differing from known biological counterparts in V1. Finally, the output layer is able to achieve a continuous representation by linearly combining the readout representation; this ultimately enables the network to (a posteriori) associate spike trains to the input movies. 

Together, these findings imply neural foundation models may be more similar in internal operation to other foundation models, rather than their biological counterparts. While this does not alter the usefulness of neural foundation models, it suggests that future architectures incorporating recurrence directly after a more light-weight retinal-like encoder and constraining feature dimensionality to match biological cell type diversity could yield models that bridge the gap between computational performance and biological plausibility.


\begin{ack}


This article has received funding from the European Commission's Marie Sk\l{}odowska-Curie Action under grant agreement no. 101207931 (LD), and by the MICIU /AEI /10.13039/501100011033 / FEDER, UE Grant no. PID2024-155187OB-I00 (LD). SWZ was supported by NIH grant 1R01EY031059 and NSF Grant 1822598. \end{ack}

\bibliographystyle{plainnat}
\bibliography{bib}

\newpage
\appendix
\renewcommand{\appendixpagename}{Appendix} 
\appendixpage

\section{Methods}
\label{app:methods}

\subsection{Online material}
Our work made use of publicly available open-source resources. Specifically, we employed the pretrained FNN model provided by \citet{wang_foundation_2025}, available at \url{https://github.com/cajal/fnn/tree/main}. For the analysis of this model, we used the stimulus generation tools and neural encoding manifold construction pipeline introduced by \citet{dyballa_population_2024}, accessible at \url{https://github.com/dyballa/NeuralEncodingManifolds}.

\subsection{FNN}
The FNN consists of five modules: perspective, modulation, \textit{encoder}, \textit{recurrent}, and \textit{readout}. The perspective and modulation modules model the mouse's state and transform the inputs to approximate the actual visual information received.
Thus, only the \textit{encoder}, \textit{recurrent}, and \textit{readout} modules perform the core computation and are the focus of this work.  

The \textit{encoder} module is a 10-layer DenseNet-style convolutional encoder. Notably, it includes 3D convolutions, which in principle enable the encoder to capture temporal patterns for up to 12 time steps. The \textit{recurrent} module is preceded by an attention layer and consists of a convolutional LSTM, followed by a single convolutional layer that produces its output.  
This feedforward–recurrent combination constitutes the core of the FNN, which is trained on all data. Finally, the \textit{readout} module is mouse-specific: it performs an interpolation on the recurrent output followed by a linear transformation to produce the FNN output. We used the FNN from session 8, scan 5. We validated results on other scans for the readout and output, and argue that our comparison at the population level is valid across animals. Moreover, the core was jointly trained on all scans.

\subsection{Input videos}\label{stimuli}
We used the visual stimuli from \citet{dyballa_population_2024}, consisting of drifting square-wave gratings and optical flows moving in eight directions. The flow stimuli include oriented (lines) and non-oriented (dots) stimuli with spatial frequencies between 0.04 and 0.5 \(\frac{\text{cycles}}{\text{deg}}\). This yields 88 unique input sequences with stochastic initial positions and velocities. The stimuli were scaled and cropped to fit the required FNN input shape of 144$\times$256 pixels. This resulted in an image sequence: 
$\{\mathbf{x}_{0}, \ldots, \mathbf{x}_{T}\}$, where each $\mathbf{x}_i \in \mathbb{R}^{H \times W}$. Stimuli were generated using the tools available at \url{https://github.com/dyballa/NeuralEncodingManifolds}.

The FNN \citep{wang_foundation_2025} processes 2.33-second sequences of 70 frames each, corresponding to 30 frames per second. Since in  \citet{dyballa_population_2024} the trials were 1.25 s long, we adapted the stimuli to contain 37 frames to maintain consistency with the FNN framework. We scaled all stimuli by a factor of 0.7 to optimize stimulus discriminability across different network layers. 

\subsection{Sampling}

Neural responses were computed using PyTorch and extracted by sampling activations from 2000 units across selected FNN layers. Within each layer, 40 feature maps were sampled. Then, 50 neurons were sampled from each feature map. Feature map sampling probabilities were calculated from the mean maximum response across all neurons within each map, while neuron sampling probabilities within each selected feature map were based on individual neuron maximum responses, biasing the sampling to include active neurons. This sampling procedure was chosen to ensure comparability to the biological results from \citet{dyballa_population_2024}. This sampling procedure was tested and validated in \citet{dyballa_population_2024}, we performed further tests with random sampling to validate this bias does not filter out relevant structures. Increasing the sampling rate beyond 2000 units did not significantly alter manifold topology but hindered cluster separation in diffusion map analysis. The resulting tensor data had dimensions \((N \times S \times O \times T)\) with \(N=2000\) neurons, \(S=11\) stimulus types, \(O=8\) orientations and \(T=37\) time steps. For manifold construction, the optimal spatial frequency was selected (resulting in \(S=6\) stimuli) whereas for classification performance all spatial frequencies were kept. We report results from a single random seed per layer, as preliminary analysis showed consistent manifold structure across different random activity samples. These neural activation tensors served as input for subsequent classification and manifold analysis. This sampling procedure was developed by \citet{dyballa_population_2024} and tested against other sampling methods there. We also experimented with the sampling procedure, finding that random sampling and increased sampling rate did not introduce qualitative changes to the manifolds.

\subsection{Stimulus adequacy}

For every FNN layer investigated in this paper, we extracted the activation to the stimulus ensemble consisting of gratings and flows (see Section~\ref{stimuli}) as well as to a 100-second-long natural input video from the MICrONS functional dataset \citep{bae_functional_2025}, downloaded from \url{s3://bossdb-open-data/iarpa_microns/minnie/functional_data/stimulus_movies/}. Both stimulus sets produced similar activation magnitudes across the entire network (see Figure~\ref{fig7}), which shows the adequacy of the stimulus ensemble used for testing the FNN. 

Additionally, we calculated the orientation selectivity (OSI) and direction selectivity (DSI) for gratings + flows and for a pink noise stimulus, as done in \citet{wang_foundation_2025}. We found comparable OSI and DSI distributions (see Figure~\ref{osi}).

\subsection{Classification accuracy}

The stimulus classification accuracy based on the individual layer activities was obtained from training multinomial logistic regression classifiers (scikit-learn, with solver L-BFGS) using 5-fold cross-validation. We used only the sampled neurons for classifying the 11 stimuli. For each layer and each time point $t$, two feature sets were constructed: (\texttt{i}) the mean activity over frames $0\rightarrow t$ (increasing window) and (\texttt{ii}) the mean activity over frames $t \rightarrow$ \textit{end} (decreasing window). The maximal classification accuracy for all feature sets is reported in Table~\ref{tab1}. For comparison, we also evaluated $K$-nearest neighbor classifiers ($K=3$) using leave-one-out cross-validation. Results are summarized in Table~\ref{tab1} and Figure~\ref{fig6}.

\subsection{Construction of decoding manifolds}

For building the \textit{decoding manifolds}, we applied PCA (scikit-learn) to the averaged activity data. In total, the \textit{decoding manifolds} contain 48 points, consisting of 6 stimuli and 8 movement directions each. The 6 stimuli were obtained from a majority vote of all neurons on the optimal spatial frequency eliciting higher responses. The decoding manifolds use different colors for each stimulus, as introduced in Figure~\ref{fig1}. Different spatial frequencies of the same stimulus are summarized with the same color. To construct \textit{decoding trajectories}, we treated each time step as a separate data point rather than averaging across time before applying PCA. We prepended each trajectory with a zero-activity time step to establish a common origin for all stimulus conditions. In both cases, we reduced the dimensionality to three components for visualization after verifying that further dimensions did not encode qualitatively new information. We constructed biological \textit{decoding trajectories} using experimental data from \citet{dyballa_population_2024}, available at \url{https://github.com/dyballa/NeuralEncodingManifolds}. For the biological decoding trajectories, we did not use the additional zero-activity time step since a baseline activity level was already provided by the inter-stimulus intervals in the experiments.

\subsection{Construction of neural encoding manifolds}

At a high level, the motivation for constructing \textit{neural encoding manifolds} is to find a space in which one can examine the global topology of neuronal populations based on their stimulus selectivities and temporal response patterns \citep{dyballa_population_2024}. The neural encoding manifold is constructed in a three-step procedure. First, a 3-tensor is built with the temporal responses from each neuron for each stimulus, and decomposed using Nonnegative Tensor Factorization (details below); each component is comprised of neural, stimulus, and temporal response factors. The neural factors then serve as position coordinates, embedding the neurons into a stimulus-response framework called the neural encoding space. Second, we construct a data graph in this neural encoding space using the IAN algorithm \citep{dyballa_ian_2023}. Third, applying diffusion maps \citep{coifman_geometric_2005, coifman_diffusion_2006} to the data graph yields the manifold. 

The methodological choices in our manifold construction procedure are made in accordance with \citet{dyballa_population_2024}, where extensive parameter analysis for biological neural data was conducted. Since \textit{neural encoding manifolds} computed with these specific parameters represent the only available comparison for biological data from the visual system, we maintained their parameter settings to ensure direct comparability between artificial and biological neural representations. We further conducted analysis for FNN-specific parameters, such as the sampling procedure, by adapting their \href{https://github.com/dyballa/NeuralEncodingManifolds/tree/main}{code} to fit the FNN requirements.

\subsubsection{Preprocessing}
The input tensor of neuronal activity (see above) was preprocessed in several steps (using NumPy and SciPy). First, the individual responses were smoothed along the time dimension using a one-dimensional Gaussian kernel with \(\sigma = 3\). Next, we grouped the stimuli into \textit{medium} versus \textit{high} spatial frequencies and selected the one exhibiting higher response magnitudes. The temporal responses for the 8 directions of motion were then concatenated together into a single vector. Finally, we normalized each response and rescaled it by the relative activations of the neuron. The resulting tensor \(\mathbf{T}\) had shape \(((N=2000) \times (S=6) \times (O * T = 296))\).

\subsubsection{Nonnegative Tensor Factorization}

Next, Nonnegative Tensor Factorization (see \citep{williams_unsupervised_2018} for an overview and applications to neuroscience) was applied to our tensor \(\mathbf{T}\). It was decomposed into typically 10--15 rank-1 tensors which are obtained from the outer product of three vectors each. The number of components was chosen separately for each data sample as specified in \citet{dyballa_population_2024}. The factors in each component are scaled to unit length, and their magnitudes absorbed by a scalar \(\lambda_r\):

\begin{align}
    \tilde{\mathbf{T}} = \sum_{r=1}^R \lambda_r \mathbf{v}_r^{(1)} \circ \mathbf{v}_r^{(2)} \circ \mathbf{v}_r^{(3)} = [\lambda; \mathbf{X}^{(1)}; \mathbf{X}^{(2)}; \mathbf{X}^{(3)}]
\end{align}

For the second equality, the factor matrices \(\mathbf{X}^{(k)}\) are constructed using the factor vectors \(\mathbf{v}_r^{(k)}\) as columns, and the vector $\lambda$ contains all individual \(\lambda_r\)s.

Decomposing the tensor \(\mathbf{T}\) into these components is an optimization problem with the following objective function and non-negativity constraints:

\begin{align}
    \operatorname*{min}_{\mathbf{X}^{(1)}, \mathbf{X}^{(2)}, \mathbf{X}^{(3)}} \frac{1}{2} || \mathbf{T} - \tilde{\mathbf{T}}||^2 \\
    \text{such that} \quad \mathbf{X}^{(k)} \geq 0, \forall k
\end{align}

The resulting decomposition is interpretable: the third group of vectors, \(\mathbf{v}_r^{(3)}\), describes different temporal response patterns; \(\mathbf{v}_r^{(2)}\) contain information about which stimuli exhibit these response patterns; and \(\mathbf{v}_r^{(1)}\) are the neuronal factors determining which neurons exhibit the response patterns characterized by \(\mathbf{v}_r^{(2)}\) and \(\mathbf{v}_r^{(3)}\). During decomposition, circular permutations were applied to detect patterns irrespective of the preferred orientations of specific neurons (again, this is necessary to ensure compatibility with the biological results from \citep{dyballa_population_2024}).

Using the OPT method from Tensor Toolbox \citep{tensor_toolbox}), we ran the decomposition 50 times (different initializations) for each number of components and dataset to ensure robust decomposition results and the choice of the number of factors, \(R\). The manifolds were robust to small changes in \(R\), therefore the heuristic for choosing \(R\) based on the explained variance of the decomposition outlined in \citet{dyballa_population_2024} proved sufficient. For building the manifolds, we used the result with smallest reconstruction error among the 50 initializations.

\subsubsection{Neural encoding space}
Following \citet{dyballa_population_2024}, we now reformulate the above decomposition to construct the neural encoding space. By defining the diagonal matrix \(\mathbf{\Lambda}\) with \(\mathbf{\Lambda}_{rr} = \lambda_r\), we obtain: 

\begin{align}
    \tilde{\mathbf{T}} = \mathbf{X}^{(1)} \mathbf{\Lambda} (\mathbf{X}^{(2)} \circ \mathbf{X}^{(3)})
\end{align}

Since the first matrix, \(\mathbf{X}^{(1)}\), represents the neuronal factors, we denote it by \(\mathcal{N}\). Now, define a matrix \(\mathbf{B}\) with columns \(\mathbf{b}_{:,r}\):

\begin{align}
    \mathbf{b}_{:,r} = vec (\mathbf{v}_r^{(2)} \circ \mathbf{v}_r^{(3)})
\end{align}

Finally, we obtain a matrix representation of \(\mathbf{T}\) with respect to neuronal factors as \(\mathbf{X}_{\mathcal{N}}\):

\begin{align}
    \mathbf{X}_{\mathcal{N}} = \mathbf{B} \mathbf{\Lambda} \mathcal{N}^T
\end{align}

This reformulation constructs the neural encoding space. The unit-norm basis vectors of this space are given by the columns of \(\mathbf{B}\). We define the neural matrix containing the positions of all neurons in this space as \(\mathcal{N}_\lambda = \mathcal{N}\mathbf{\Lambda}\). The distances between any two neurons in this space reflect their similarity in stimulus-selective temporal response patterns. Intuitively, neurons with similar selectivity profiles and temporal dynamics should be positioned close together, while neurons with dissimilar response characteristics should be farther apart. 

\subsubsection{Iterated adaptive neighborhoods (IAN)}
Within this neural encoding space, we construct a weighted graph of the data by inferring a similarity kernel. This is achieved using the Iterated Adaptive Neighborhoods (IAN) algorithm \citep{dyballa_ian_2023}, which infers an adaptive local kernel without the need for pre-specifying a fixed neighborhood size.

IAN first constructs the unweighted Gabriel graph for the data points. In addition, a weighted graph is constructed using a multiscale Gaussian kernel based on the discrete neighborhood graphs. Subsequently, the graph is iteratively pruned by ensuring consistency between the discrete and continuous neighborhoods. The resulting weighted graph is represented by the adjacency (kernel) matrix \(\mathbf{K}\). This matrix contains similarities computed using locally tuned Gaussian kernels.

\subsubsection{Diffusion maps}
Diffusion Maps \citep{coifman_geometric_2005, coifman_diffusion_2006} are a dimensionality reduction technique that retain distances and preserve the intrinsic geometry of the manifold. The diffusion process is based on graph Laplacian normalization from spectral graph theory. 

In detail, we use the weighted graph obtained from IAN as the weighted adjacency matrix \(\mathbf{K}\). The first step is to normalize and symmetrize it to produce \(\mathbf{M}_s\):

\begin{align}
    \mathbf{d}_i &= \sqrt{\sum_j \mathbf{K}_{ij} + \epsilon} \\
    \mathbf{M}_s &= \frac{\mathbf{K}}{\mathbf{d} \mathbf{d}^T}
\end{align}
    
This normalization ensures that nodes of high degree do not dominate the analysis. We then calculate the spectral decomposition of \(\mathbf{M}_s\) with eigenvalues \(\lambda_0 = 1 \geq \lambda_1 \geq \lambda_2 ...\) and eigenvectors \(\boldsymbol{\psi}_i\) for \(t=1\) diffusion steps using \(L=20\) eigenvalues: 

\begin{align}
    \mathbf{M}_{s, ij}^t = \sum_{l = 0}^L \lambda_l^{2t} \boldsymbol{\psi}_l(i) \boldsymbol{\psi}_l(j)
\end{align}

Finally, from the spectral decomposition, we obtain the diffusion map with diffusion coordinates:

\begin{align}
    \Psi_t(i) =
    \begin{pmatrix}
    \lambda_0^t \boldsymbol{\psi}_0(i)  \\
    \lambda_1^t \boldsymbol{\psi}_1(i)  \\
    \vdots  \\
    \lambda_{L-1}^t \boldsymbol{\psi}_{L-1}(i)
    \end{pmatrix}
\end{align}

Plotting the data using these diffusion coordinates yields the \textit{neural encoding manifold}. 

\subsubsection{Encoding manifold visualization}

For visualization purposes, we optionally applied metric multidimensional scaling (MDS) to the diffusion map coordinates. This was done by computing pairwise squared Euclidean distances using the first diffusion coordinates, constructing the corresponding Gram matrix \(\mathbf{G} = -0.5 * \mathbf{D}^2\), and applying kernel PCA to obtain a lower-dimensional embedding. This preserves the distance relationships from the diffusion map while combining multiple diffusion coordinates, enabling a clearer visualization of the manifold structure.

Based on the manifold topology, we selected groups of neurons to investigate via their PeriStimulus Time Histograms (PSTH). We averaged their activity across trials and constructed the PSTHs as a 2-D heatmap, where each row contains the temporal activity in response to a particular direction of motion (as displayed in Figure~\ref{fig1}). Additionally, we calculated the average response intensity over time for these groups and reported the s.e.m. using the shaded regions (see insets in Figure~\ref{fig2}A,D).

\subsection{Manifold metrics}

We computed the following metrics to analyze neural encoding manifolds in Figures~\ref{fig4} and \ref{fig5}:

\paragraph{OSI} An Orientation Selectivity Index was computed as:

\begin{align}
    OSI_n &= \max_s OSI_{n,s}\\
    OSI_{n,s} &= \frac{R_{n,s}(\theta^*) - \tfrac{1}{2} \big(R_{n,s}(\theta^* + 90^{\circ}) + R_{n,s}(\theta^* - 90^{\circ})\big)}{R_{n,s}(\theta^*) + \tfrac{1}{2} \big(R_{n,s}(\theta^* + 90^{\circ}) + R_{n,s}(\theta^* - 90^{\circ})\big) + \epsilon}\\
    \theta^* &= \argmax_\theta R_{n,s}(\theta)
\end{align}

where \(R_{n,s}(\theta)\) is the mean response of neuron \(n\) to stimulus \(s\) at orientation \(\theta\).  

\paragraph{Mean activity} We computed mean activities by averaging each neuron's response across all time steps, directions, and stimuli.

\paragraph{Temporal variance} We calculated the temporal variance for each stimulus and direction combination. We then averaged these variances for each neuron.

\paragraph{Preferred stimulus} The preferred stimulus for each neuron was obtained by finding the stimulus exhibiting the highest average activity across stimuli and time steps for each neuron.

\subsubsection{Tubularity of a Bundle of Trajectories}\label{tubular:methods}

\textbf{Goal.} To investigate neural dynamics, we modeled trajectories by bundling them into tubular neighborhoods around a central skeleton \citep{budanur2023predictioncontrolspatiotemporalchaos}. We operationlized this idea for discrete data using the tubular neighborhood theorem \citep{da_silva_lectures_2008}, which guarantees that smooth submanifolds admit non-intersecting neighborhoods diffeomorphic to their normal bundles. Let $\{\gamma_i\}_{i=1}^m \subset \mathbb{R}^D$ denote a set of $m$ trajectories (curves). We define this set as \emph{tubular} if it remains close to a common centerline $c$ and exhibits minimal transverse intersections. Formally, the tube is obtained by expanding $c$ with a radius profile $R(\cdot)$ such that all points at parameter $u$ within distance $R(u)$ of $c(u)$ are included. In practice, curves are first clustered (e.g., via HDBSCAN \citep{hdbscan} using the Sobolev $H^1$ metric, or with ground truth) to separate distinct tubes before computing tubularity scores. We introduce \textit{tightness}, which measures how tight a group of curves is around the centerline, and \textit{crossings}, measuring how many transverse crossings occur in each trajectory bundle.

\subsection{Tubularity}
Before calculating tubularity metrics, we standard-scale the data and apply PCA to obtain a 10-dimensional embedding, thereby speeding up the computation. While visualizations use only the first 2-3 dimensions, all metrics are calculated in the 10-dimensional space. To ensure comparability, we resampled all trajectories to length 100. For statistical analysis, we generated 100 bootstrapped samples, and using ground-truth clusters, performed Bonferroni-corrected Mann-Whitney U tests on our hypotheses.

We formalize how ``tight'' a group of curves is around the centerline: We reparameterize each curve by normalized arc length $u\!\in\![0,1]$ and resample to $\{u_k\}_{k=1}^M$. Let $x_i(u_k)\!\in\!\mathbb{R}^D$ denote the samples and $\tau_i(u_k)$ their unit tangents. We define the \emph{mean curve} as the pointwise average:

$$
c(u_k)\;=\;\frac{1}{m}\sum_{i=1}^m x_i(u_k),\qquad
r_i(u_k)\;=\;\|x_i(u_k)-c(u_k)\|.
$$

The tightness score is calculated by averaging quantile tube radii across bins $\{I_b\}_{b=1}^B$ that partition $[0,1]$, using a high quantile $q \in [0.8, 0.95]$ to ensure robustness to noise. We normalize each tube's tightness score by tube length.

$$
S_{\mathrm{tight}}
\;=\;
\frac{1}{B}\sum_{b=1}^{B}\operatorname{quantile}_q\{\,r_i(u):u\in I_b\ \text{over all curves}\,\}.
$$

The second quantity assessed is the uniformity of the tubes relative to one another. That is, the degree to which crossings occur in our defined bundle of curves. Tubes are considered disorganized when distinct curves pass near each other with \emph{transverse} directions. Let
$d_{ij}(u,v)=\|x_i(u)-x_j(v)\|$ and
$\phi_{ij}(u,v)=1-\langle \tau_i(u),\tau_j(v)\rangle^2\in[0,1]$ (large for near-orthogonal tangents). Using a Gaussian kernel $K_\varepsilon(\rho)=\exp(-\rho^2/(2\varepsilon^2))$, we softly count encounters:

$$
\mathcal{X}_\varepsilon
=\frac{2}{m(m-1)}
\sum_{i<j}
\int_0^1\!\!\int_0^1
K_\varepsilon\!\big(d_{ij}(u,v)\big)\,
\phi_{ij}(u,v)\,du\,dv.
$$

$S_{\mathrm{tight}}$ and $S_{\mathrm{cross}}$ only depend on distance, unit-tangent inner product, and arc-length. Therefore, they are invariant to translations, rotations, and re-timing. We emphasize that, for both scores, smaller values indicate more tubular curve bundles, while larger values indicate fewer tubular curve bundles.

\subsection{Visualizations}
Interactive three-dimensional plots of the manifolds were computed using Plotly. Other plots were created with Matplotlib and TUEplots.

\subsection{Software}
\label{software}
All software (Table~\ref{tab2}) is used in accordance with its respective license.

\begin{table}[h!]
\centering
\caption{Software packages used in this work. }
\label{tab2}
\begin{tabular}{lll}
\toprule
\textbf{Package} & \textbf{Version} & \textbf{License} \\
\midrule
MATLAB Tensor Toolbox \citep{tensor_toolbox} & 3.6     & BSD-2 \\
IAN \citep{dyballa_ian_2023}                 & 1.1.2   & BSD-3 \\
NeuralEncodingManifolds \citep{dyballa_population_2024} & N/A & BSD-2\\
NumPy \citep{numpy}                          & 1.25.0  & BSD-3 \\
SciPy \citep{SciPy}                          & 1.15.3  & BSD-3 \\
scikit-learn \citep{scikit-learn}            & 1.7.1   & BSD-3 \\
PyTorch \citep{pytorch}                      & 2.6.0   & MIT \\
Matplotlib \citep{matplotlib}                & 3.10.1  & PSF-based (BSD-compatible) \\
Plotly \citep{plotly}                        & 6.0.0   & MIT \\
TUEplots \citep{tueplots}                    & 0.2.0   & MIT \\
\bottomrule
\end{tabular}
\end{table}

\subsection{Compute}
\label{compute}

The experiments were conducted on an HPC cluster. FNN sampling uses randomly selected GPUs (RTX 2080 Ti, or better). All other experiments were performed on CPU. All experiments required less than 30 GB memory. In total, 10 tensor decomposition experiments were run on CPU, each taking 2 days on a single CPU. Preliminary results not included in the paper required another 50 tensor decomposition experiments.

\section{Data and code availability}

The code is available at \url{https://github.com/JohannesBertram/FNN_Manifolds}

\newpage
\section{Supplemental Figures}


\begin{figure}[h]
  \centering
  \includegraphics[width=\textwidth]{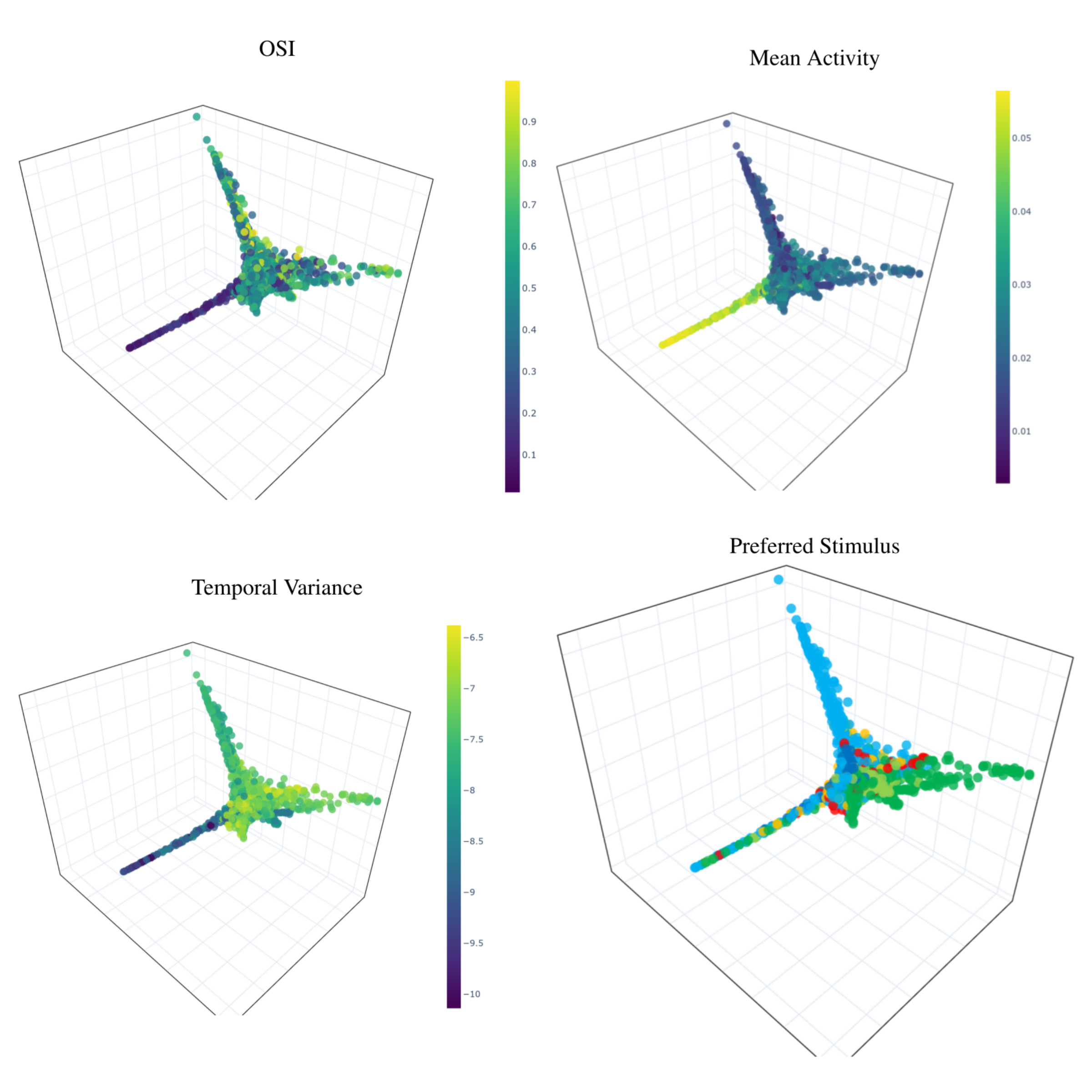}
  \caption{\textbf{Feedforward encoder L8}: Supporting information for \hyperref[fig2]{Figure 2}. The intensity arm is clearly visible, exhibiting high mean activity. The low temporal variance, low Orientation Selectivity Indices, and unstructured preferred stimuli show the absence of complex activity patterns in this intensity arm.
  }
  \label{fig4}
\end{figure}

\begin{figure}[h]
  \centering
  \includegraphics[width=\textwidth]{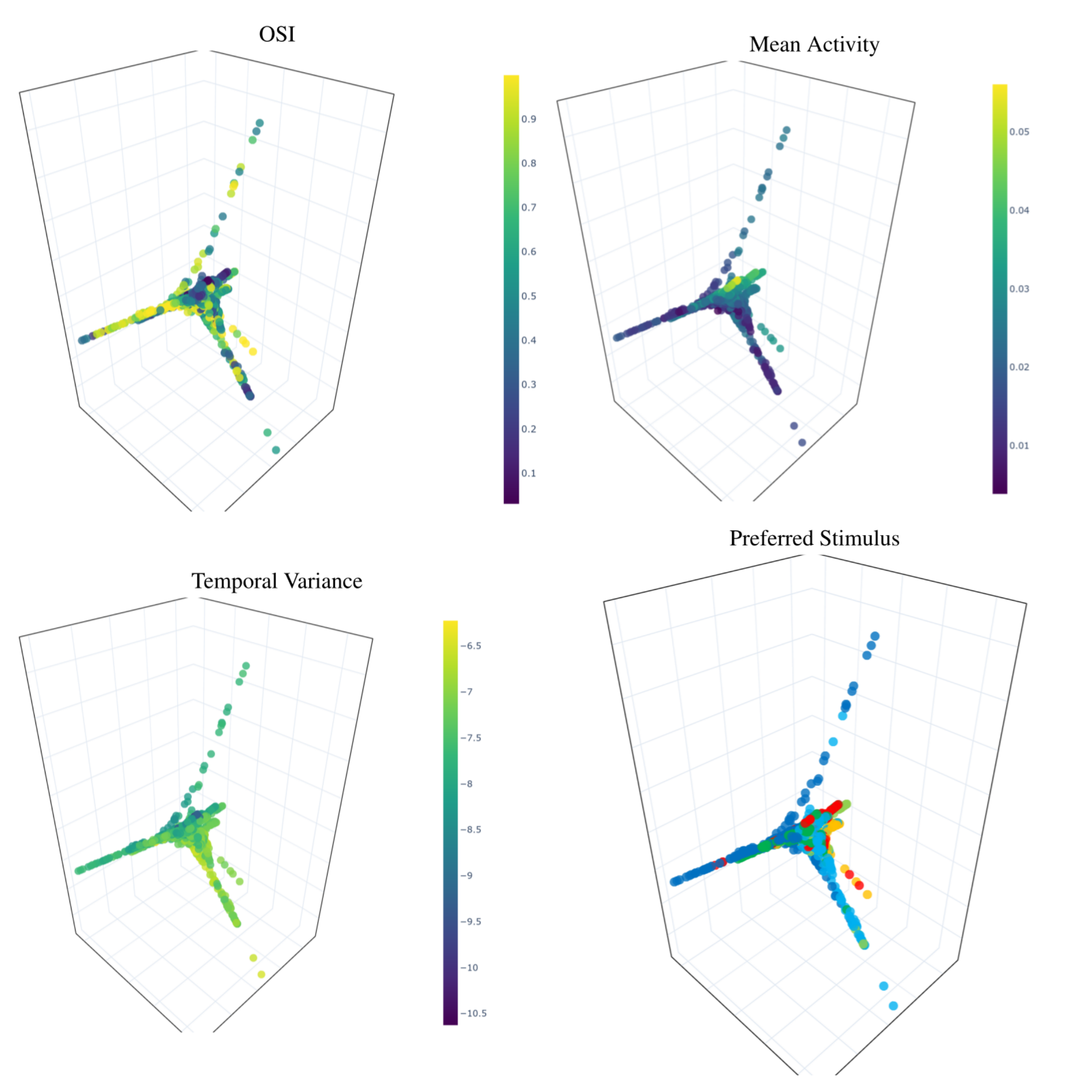}
  \caption{\textbf{Recurrent hidden state}: Supporting information for \hyperref[fig2]{Figure 2}. In contrast to \hyperref[fig4]{Figure 4}, there is no intensity arm dominating the manifold structure. Instead, all arms show structured, complex selectivity patterns.
  }
  \label{fig5}
\end{figure}

\begin{figure}[h]
  \centering
  \includegraphics[width=\textwidth]{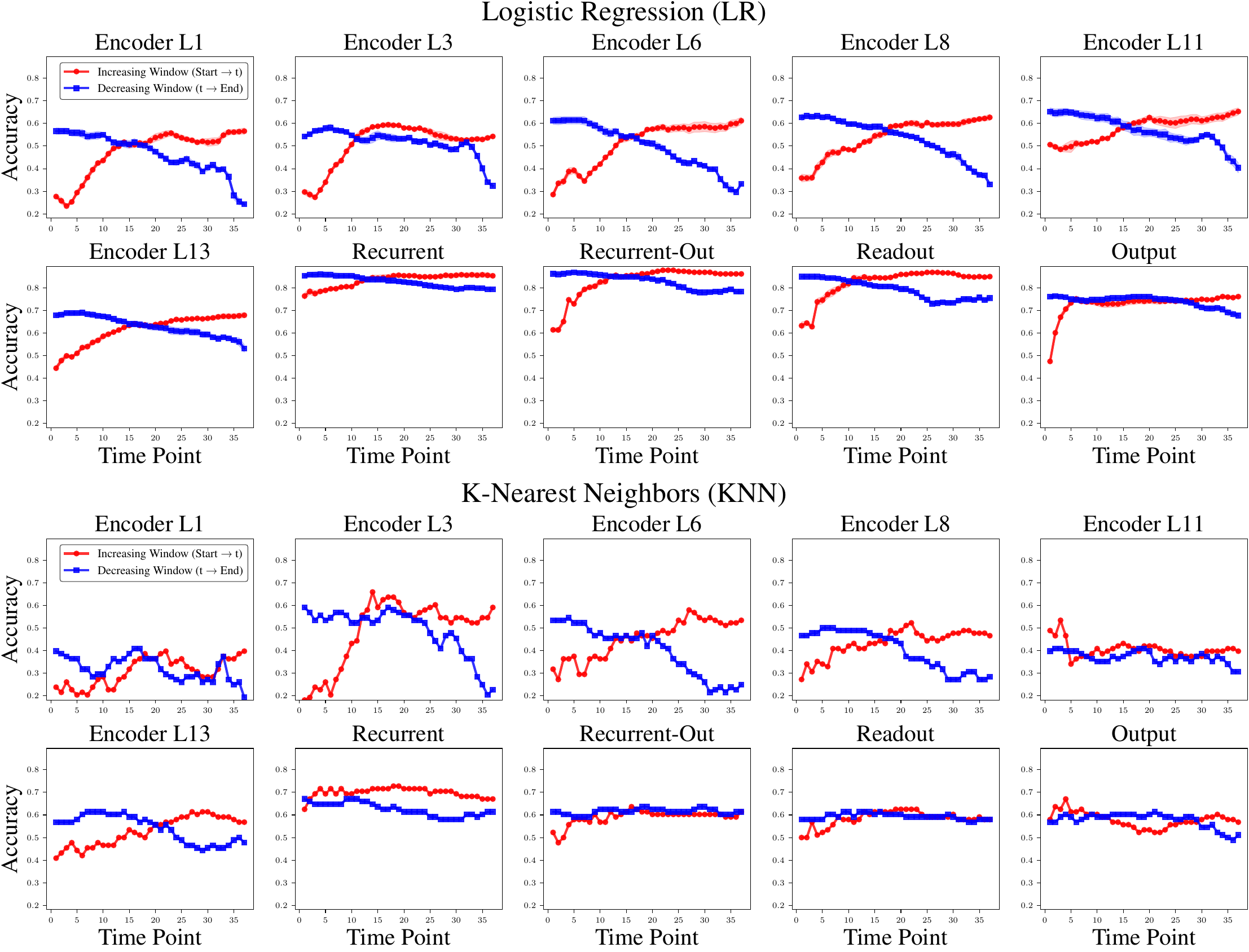}
  \caption{\textbf{Logistic regression (LR, top) and K-Nearest Neighbor (KNN, K=3, bottom) classifier accuracy} for each layer. We use increasing time windows (timesteps 0 \(\rightarrow\) t, red) or decreasing time windows (t \(\rightarrow\) 37, blue) to calculate the accuracies. Shaded regions for LR show the s.e.m. The maxima across panels are summarized in Table~\ref{tab1}.
  }
  \label{fig6}
\end{figure}

\begin{figure}[h]
  \centering
  \includegraphics[width=\textwidth]{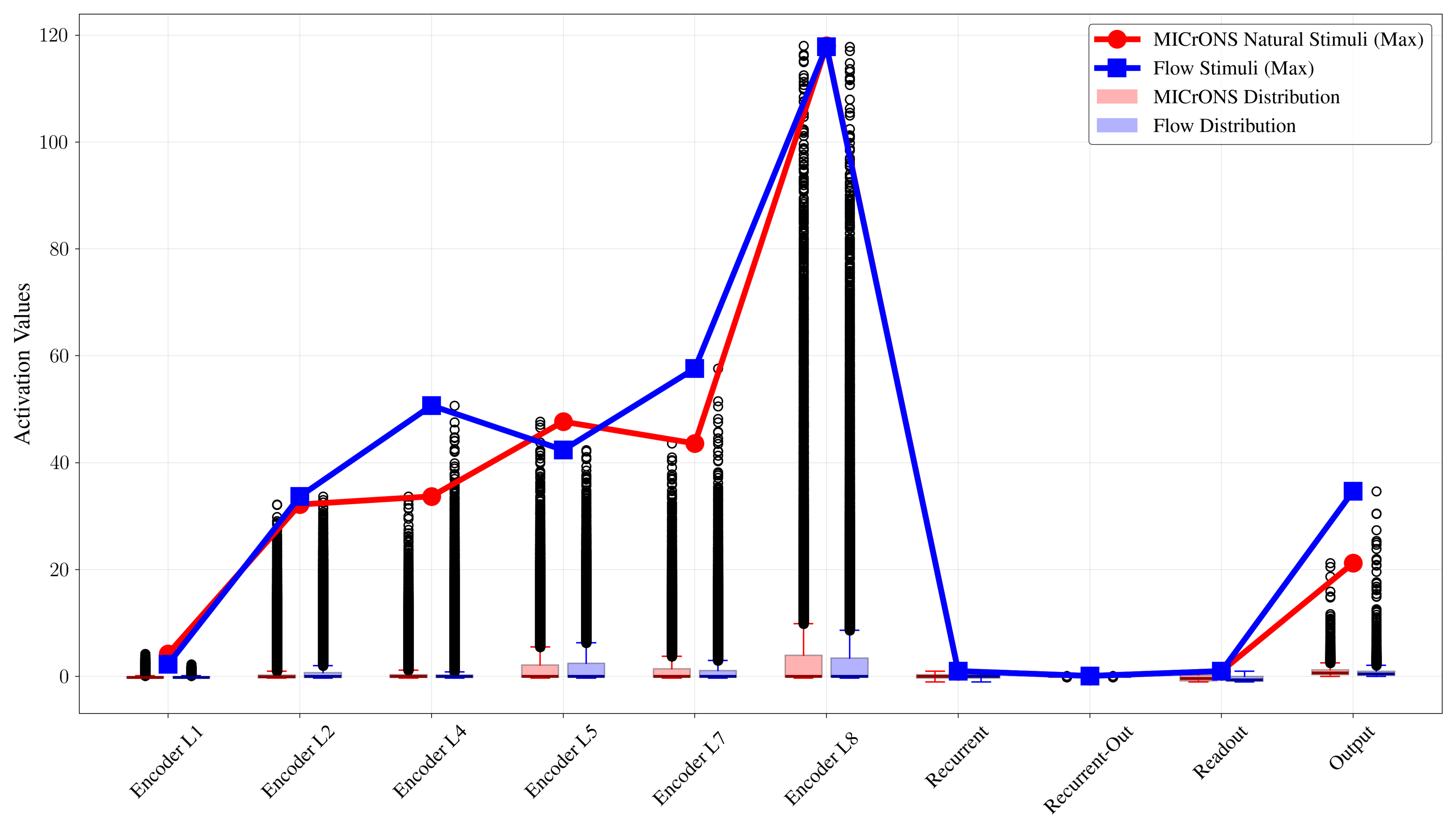}
  \caption{\textbf{Activation function output distributions and maxima} for natural MICrONS \citep{bae_functional_2025} input videos and the flow stimulus ensemble \citep{dyballa_population_2024}. The comparable activity across network layers shows the adequacy of investigating the FNN with flow stimuli. The differences in magnitudes across layers are explained by the activations functions (GELU in the \textit{encoder}, Tanh in the \textit{recurrent} and \textit{readout} modules).
  }
  \label{fig7}
\end{figure}

\begin{figure}[h]
    \centering
    \includegraphics[width=\linewidth]{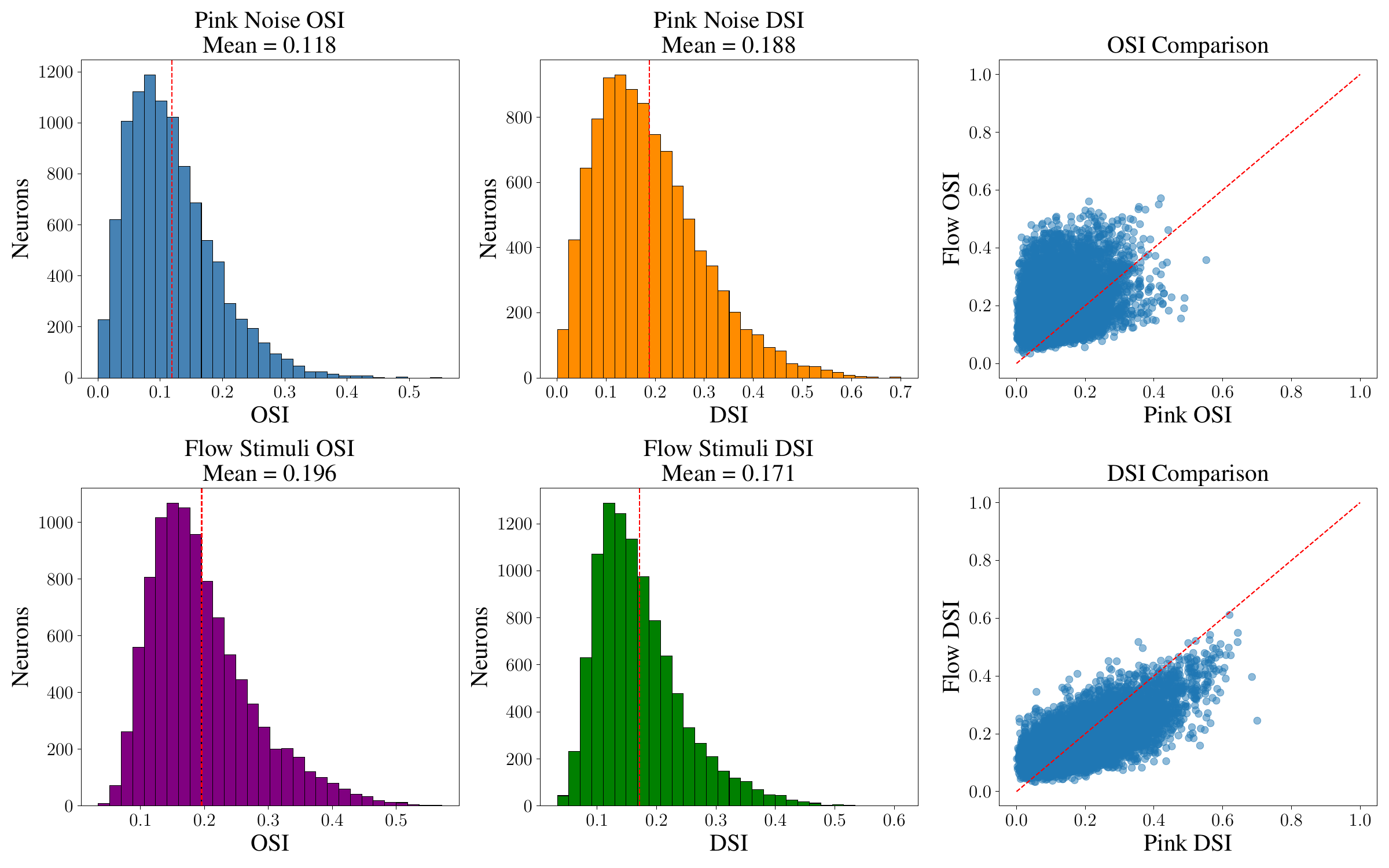}
    \caption{\textbf{OSI and DSI of FNN output} for pink noise (as used in \citet{wang_foundation_2025}) and for the stimulus ensemble from \citet{dyballa_population_2024}, meaned over the different stimuli.}
    \label{osi}
\end{figure}

\begin{table}[b]
\vspace{-2mm}
\caption{\textbf{Tubularity metrics for biological and FNN data}. Low tightness and crossings values indicate high tubularity. Aligning with the visualization in Figure~\ref{fig:tubes}, the biological trajectories show highly tubular organizations compared to FNN. Method details in Appendix~\ref{tubular:methods}.}
\label{tab:tubularity}
\centering

\begin{tabular}{cccccc}
\toprule
\textbf{Layer} & \multicolumn{2}{c}{\textbf{Ground Truth Labels}} & \multicolumn{3}{c}{\textbf{HDBSCAN Labels}} \\
\midrule
& $\mathbf{S_{\text{tight}}}$ & $\mathbf{S_{\text{cross}}}$ & $\mathbf{S_{\text{tight}}}$ & $\mathbf{S_{\text{cross}}}$ & \textbf{Clusters} \\
\midrule
Retina & $0.0688$ & $1.29 \times 10^{-05}$ & $0.1017$ & $1.06 \times 10^{-05}$ & 4 \\

V1 & $0.1357$ & $4.06 \times 10^{-05}$ & $0.1859$ & $3.50 \times 10^{-05}$ & 4 \\

Enc1 & $0.2018$ & $2.87 \times 10^{-04}$ & $0.7680$ & $1.66 \times 10^{-04}$ & 1 \\

Enc13 & $1.9885$ & $1.77 \times 10^{-06}$ & $4.3461$ & $1.09 \times 10^{-06}$ & 3 \\

Rec & $0.1228$ & $2.65 \times 10^{-07}$ & $0.1697$ & $1.53 \times 10^{-07}$ & 4 \\

RecOut & $0.1209$ & $5.72 \times 10^{-07}$ & $0.1650$ & $5.34 \times 10^{-07}$ & 4 \\

Readout & $0.3307$ & $3.96 \times 10^{-06}$ & $0.4320$ & $5.40 \times 10^{-06}$ & 4 \\

Output & $0.1483$ & $3.53 \times 10^{-06}$ & $0.2784$ & $1.12 \times 10^{-06}$ & 3 \\
\bottomrule
\end{tabular}
\end{table}






\begin{figure}[h]
    \centering
    \includegraphics[width=\linewidth]{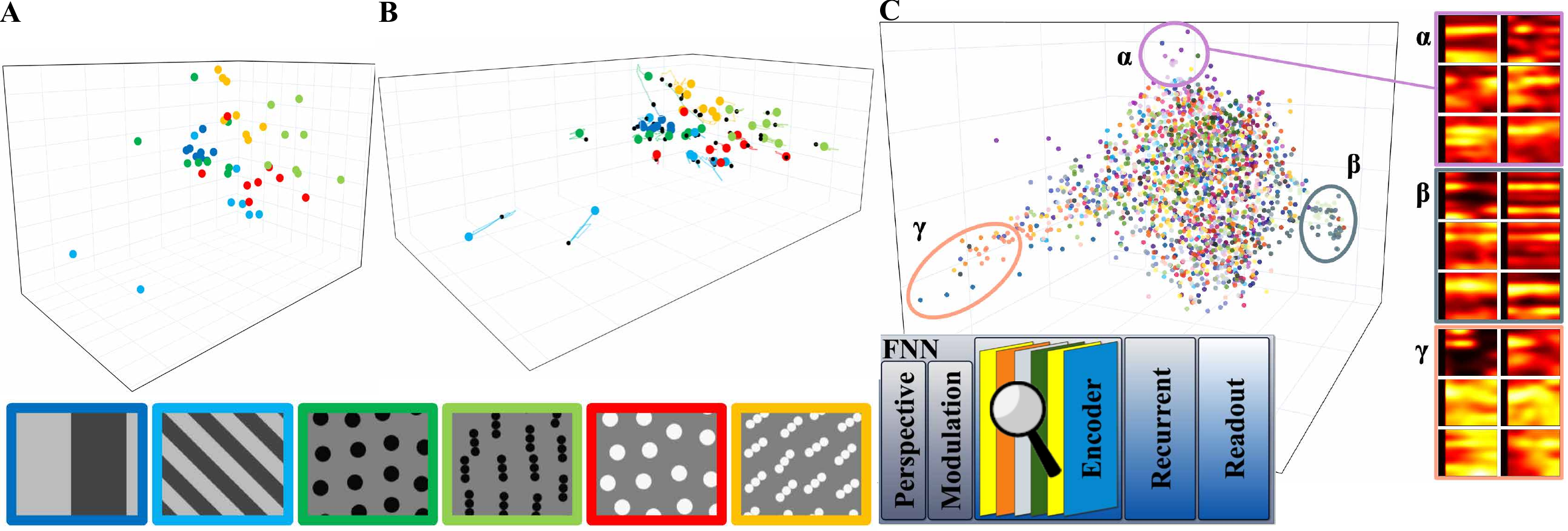}
    \caption{\textbf{Encoder L8 decoding manifold, trajectories and encoding manifold without intensity artifacts}. Without the intensity artifacts there is no temporal development at all in the decoding trajectories (comparable to encoder L1) apart from the jump after the 0-th step. The non-selective high intensity neurons are padding artifacts at the edges of the image. In the encoder, due to spatial convolutions, the effect of these artifacts spreads out across the feature maps. This is supported by the intensity smoothly organizing the manifold with a transition from intensity-only neurons to selective responses. In the recurrent stage, the function of the attention layer is capable of filtering exactly those artifacts out. The artifacts are reintroduced by the recurrent-output convolution, but then filtered out by the readout interpolation from central neurons only.}
    \label{fig:no_intensity}
\end{figure}

\end{document}